\documentclass[12pt,a4paper]{article}

\usepackage{amsmath} \usepackage{amsfonts, amssymb}
\usepackage{amscd}

\newcommand{\EQ}{\begin{equation}}
\newcommand{\EN}{\end{equation}}

\newtheorem{theo}{Theorem}

\newtheorem{prop}{Proposition}
\newtheorem{lemm}{Lemma}

\newtheorem{ex}{Example}
 \newcommand{\pr}{\indent{\em Proof: \ }}
\newcommand{\qed}{\hspace*{5 mm}$\Box$\bigskip}

\newcommand{\scirc}{\mbox{\footnotesize$\circ\;$}}
\newcommand{\Z}{{\mathbb{Z}}}

\newcommand{\F}{{\mathbb{F}}}

\newcommand{\dd}{\displaystyle}

\newcommand{\codi}{{\cal C}}
\newcommand{\D}{{\cal D}}

\newcommand{\add}{\Z_2\Z_4}

\newenvironment{demo}{\noindent {\pr}\ }{\qed}
\newcommand{\bgamma}{\bar{\gamma}}
\newcommand{\bdelta}{\bar{\delta}}

\title{On the intersection of additive
perfect codes \thanks{This work has been partially supported by
the Spanish MEC and the European FEDER Grant MTM2006-03250 and
also by the UAB grant PNL2006-13.}}

\author{J. Rif{\`a},$\hspace*{-1.4mm}^\ddag$ F. Solov'eva\thanks{F. Solov'eva  is
with the Sobolev Institute of Mathematics, Novosibirsk, Russia.
(email:~sol@math.nsc.ru)}, M. Villanueva\thanks{J. Rif{\`a} and M.
Villanueva are with the Department of Information and
Communications Engineering,
                             Universitat Aut{\`o}noma de Barcelona,
                             08193-Bellaterra, Spain.
                             (email:~\{josep.rifa,
                             merce.villanueva\}@autonoma.edu)}%
}

\date{November 13, 2006} \begin{document}

\maketitle  \begin{abstract} The intersection problem for additive
(extended and non-extended) perfect codes, i.e. which are the
possibilities for the number of codewords in the intersection of
two additive codes $\codi_1$ and $\codi_2$ of the same length, is
investigated. Lower and upper bounds for the intersection number
are computed and, for any value between these bounds, codes which
have this given intersection value are constructed.

For all these codes $\codi_1$ and $\codi_2$, the abelian group
structure of the intersection codes $\codi_1 \cap \codi_2$ is
characterized. The parameters of this abelian group structure
corresponding to the intersection codes are computed and lower and
upper bounds for these parameters are established. Finally,
constructions of codes the intersection of which fits any
parameters between these bounds are given.


\end{abstract}

\section{Introduction and basic definitions}

Let $\F^n$ be an $n$-dimensional vector  space over the finite
field $\Z_2$. The \textit{Hamming distance} $d(v,s)$ between two
vectors $v,s \in \F^n$ is the number of coordinates in which $v$
and $s$ differ.

A \textit{binary code} $C$ of length $n$ is a subset of $\F^n$.
The elements of a code are called {\em codewords}. The
\textit{minimum distance} $d$ of a code $C$ is the minimum value
of $d(a,b)$, where $a,b \in C$ and $a \neq b$. The {\em error
correcting capability} of a code $C$ is the value $e=\lfloor
\frac{d-1}{2} \rfloor$ and $C$ is called an $e$-error correcting
code. Two binary codes $C_1$ and $C_2$ of length $n$ are {\em
isomorphic} if there exists a coordinate  permutation $\pi$ such
that $C_2=\{ \pi(c) \ | \ c\in C_1 \}$. They are {\em equivalent}
if there exist a vector $a\in \F^n$ and a coordinate permutation
$\pi$ such that $C_2=\{ a+\pi(c) \ | \ c\in C_1 \}$.

A {\em binary perfect 1-error correcting code} (briefly in this
paper, {\em binary perfect code}) $C$ of length $n$ is a subset of
$\F^n$, with minimum distance $d=3$, such that all the vectors in
$\F^n$ are within distance one from a codeword. For any $t>1$
there exists exactly one binary linear perfect code of length
$2^t-1$, up to equivalence, which is the well-known {\it Hamming
code}. An {\it extended code} of the code $C$ is a code resulting
from adding an overall parity check digit to each codeword of $C$.

\medskip The intersection problem for binary perfect codes was
proposed by Etzion and Vardy in \cite{EV98}.  They presented  a
complete
 solution of the intersection problem for binary Hamming
codes: for each $t\geq 3$, there exist two Hamming codes $H_1$ and
$H_2$ of length $n=2^t-1$ such that the number of codewords
$\eta(H_1, H_2)$ in the intersection of these two codes is
\begin{equation}\label{EtzionVardyHamIntersection}\eta(H_1,
H_2)=2^{n-r}\quad \hbox{for}\quad r=t, t+1,
\ldots,2t.\end{equation} They found the smallest intersection
number for binary perfect codes of any admissible length consists
of two codewords and investigated the intersection problem for
binary perfect codes given by switchings of the binary Hamming
codes. Last result was improved for binary perfect codes in
\cite{AHS05,AHS06} using switching approach. Bar-Yahalom and
Etzion  solved the intersection problem for $q$-ary cyclic codes
in \cite{BYE97}. The intersection problem for $q$-ary perfect
codes is investigated in \cite{SL06}. In \cite{PV06}, the
intersection problem is also solved  for Hadamard codes of length
$2^t$ and of length $2^ts$ ($s$ odd and $t\geq 6$), as long as
there exists a Hadamard matrix of length $4s$.

\medskip The present paper is structured in the following way.
This section contains the basic definitions about additive codes
and their duals. These codes after a Gray map lead to the
$\Z_4$-linear and $\add$-linear codes. We present some useful
properties to manage those which are perfect codes and  extended
perfect codes. Section \ref{IntersectionQuaternaryCodes} is
devoted to the generic class of additive codes establishing some
results about the abelian group structure of the intersection and
the intersection numbers for these codes. In Section \ref{qlpc} we
settle the intersection problem for the additive extended perfect
codes with $\alpha=0$, i.e. for the quaternary linear perfect
codes. We establish the lower and upper bounds for the
intersection number and also for the parameters of the abelian
group structure of the intersection of these codes. Moreover, we
prove the existence of codes with all the allowed parameters
between these bounds. Finally, Section \ref{aepc} reaches the same
results than Section \ref{qlpc} but now for additive extended
perfect codes with $\alpha\not=0$.

\subsection{Additive codes} Let $\Z_2$ and $\Z_4$ be the ring of
integers modulo 2 and modulo 4, respectively. Let $\F^n$ be the
set of all binary vectors of length $n$ and let $\Z_4^n$ be the
set of all quaternary vectors of length $n$. As we said before,
any non-empty subset $C$ of $\F^n$ is a binary code and a subgroup
of $\F^n$ is called a {\it binary linear code} or a {\it
$\Z_2$-linear code}. Equivalently, any non-empty subset ${\cal C}$
of $\Z_4^n$ is a quaternary code and a subgroup of $\Z_4^n$ is
called a {\it quaternary linear code}.

Let $\codi$ be a subgroup of $\Z_2^{\alpha}\times\Z_4^{\beta}$ and
let $C=\Phi(\codi)$, where $\Phi: \Z_2^{\alpha}\times\Z_4^{\beta}
\longrightarrow \Z_2^{n}$, $n=\alpha+2\beta$, is given by
$\Phi(x,y)=(x,\phi(y))$ for any $x$ from $\Z_2^\alpha$ and any $y$
from $\Z_4^\beta,$ where
$\phi:\Z_4^\beta\;\longrightarrow\;\Z_2^{2\beta}$ is the usual
Gray map, that is, $\phi(y_1,\ldots,y_\beta)=
(\varphi(y_1),\ldots,\varphi(y_\beta)),$ and
$\varphi(0)=(0,0),\varphi(1)=(0,1),\varphi(2)=(1,1)$,
$\varphi(3)=(1,0)$.

Since $\codi$ is a subgroup of  $\Z_2^{\alpha}\times
\Z_4^{\beta}$, it is also isomorphic to an abelian structure like
$\Z_2^{\gamma}\times \Z_4^{\delta}$. Therefore, we have that
$|\codi|=2^\gamma 4^\delta $ and the number of order two codewords
in $\codi$ is $2^{\gamma+\delta}$. We call such code $\codi$ an
{\it additive code of type $(\alpha,\beta;\gamma,\delta)$} and the
binary image $C=\Phi(\codi)$ a {\it $\Z_2\Z_4$-linear code of type
$(\alpha,\beta;\gamma,\delta)$}. In the specific case $\alpha=0$
we see that $\codi$ is a quaternary linear code and the code $C$
is called a {\it $\Z_4$-linear code}. Note that the length of the
binary code $C=\Phi(\codi)$ is $n=\alpha+2\beta$.

Moreover, although $\codi$ could not have a basis, it is
interesting and adequate to define a generator matrix for $\codi$
as:
$$
   {\cal G}= \left ( \begin{array}{c|c}
        B_2 & Q_2 \\
        \hline
        B_1 & Q_1 \\
    \end{array}\right ),
$$ where $B_2$ is a $\gamma\times \alpha$ matrix; $Q_2$ is a
$\gamma \times \beta$ matrix; $B_1$ is a $\delta\times \alpha$
matrix and $Q_1$ is a $\delta\times \beta$ matrix. Matrices $B_1,
B_2$ are binary and $Q_1, Q_2$ are quaternary, but the entries in
$Q_2$ are only zeroes or twos. In what follows we denote the
additive span of the union of two additive codes $\codi_1$ and
$\codi_2$ by $\langle \codi_1,\codi_2 \rangle.$

Notice that in some cases we will need to refer the type of an
additive code as $(\alpha,\beta;\gamma,\delta; \kappa)$, where
$\kappa$ comes from the following consideration. Let $X$
(respectively $Y$) be the set of $\Z_2$ (respectively $\Z_4$)
coordinate positions. Call $\codi_X$ (respectively $\codi_Y$) the
code $\codi$ restricted to the $X$ (respectively $Y$) coordinates.
Let $\D$ be the subcode of $\codi$ which contains all order two
codewords and let $\kappa$ be the dimension of $\D_X$, which is a
binary linear code. For the case $\alpha=0$, we will write
$\kappa=0$.

Two additive codes $\codi_1$ and $\codi_2$ both of the same length
are said to be {\it equivalent}, if one can be obtained from the
other by permuting the coordinates and changing the signs of
certain coordinates. Additive codes which differ only by a
permutation of coordinates are said to be {\it isomorphic}.

\subsection{Duality of additive codes}

We will use the following definition (see~\cite{RP97}) of the
inner product in $\Z_2^{\alpha}\times \Z_4^{\beta}$: \EQ
\label{inner}
  \langle u,v \rangle=2(\sum_{i=1}^{\alpha}
  u_iv_i)+\sum_{j=\alpha+1}^{\alpha+\beta}
u_jv_j\in \Z_4, \EN
 where $u,v\in \Z_2^{\alpha}\times
\Z_4^{\beta}$. Note that when $\alpha=0$ the inner product is the
usual one for $\Z_4$-vectors (i.e. vectors over $\Z_4$) and when
$\beta=0$ it is twice the usual one for $\Z_2$-vectors.

We can also write \EQ\label{inner2}\langle u,v \rangle=
u{\cdot}J_n{\cdot}v^T,\EN where
$\dd J_n=\left (\begin{array}{c|c} 2I_{\alpha}&0\\
\hline 0&I_{\beta}\end{array}\right )$ is a diagonal quaternary
matrix.

The {\it additive dual code} of $\codi$, denoted by ${\cal
C}^\perp$, is defined in the standard way $${\cal C}^\perp=\{u\in
\Z_2^\alpha \times \Z_4^\beta \;|\; \langle u,v \rangle =0 \mbox{
for all } v\in {\cal C}\}.$$ The corresponding binary code
$\Phi({\cal C}^\perp)$ is denoted by $C_\perp$ and called the {\it
$\Z_2\Z_4$-dual code} of $C$. In the case $\alpha=0$, ${\cal
C}^\perp$ is also called the {\it quaternary dual code} of ${\cal
C}$ and $C_\perp$ the {\it $\Z_4$-dual code} of $C$.

The additive dual code $\mathcal{C}^\perp$ is also an additive
code, that is a subgroup of $\Z_2^{\alpha}\times \Z_4^{\beta}$.
Its weight enumerator polynomial is related to the weight
enumerator polynomial of $\mathcal{C}$ by McWilliams identity (see
\cite{Del73}).  Notice that $C$ and $C_\perp$ are not dual in the
binary linear sense but the weight enumerator polynomial of
$C_\perp$ is the McWilliams transform of the weight enumerator
polynomial of $C$. So, we have (see \cite{RP97})
\EQ\label{McWilltransform}
|\mathcal{C}||\mathcal{C}^\perp|=2^{\alpha+2\beta}. \EN

It is known (see \cite{BF06}) that the additive dual code of an
additive code  $\codi$ of type
($\alpha,\beta;\gamma,\delta;\kappa$), denoted by ${\cal
C}^\perp$, is of type ($\alpha,\beta;\gamma',\delta';\kappa'$),
where
\EQ\label{parameters} \begin{split}\gamma' &= \alpha + \gamma - 2\kappa,\\
\delta' &= \beta - \gamma - \delta + \kappa,\\
 \kappa'&=\alpha-\kappa.\end{split}
\EN Notice that given a quaternary linear code $\codi$ of type
($0,\beta;\gamma,\delta$), the additive dual code ${\cal C}^\perp$
is of type (see \cite{Sole})
\EQ\label{quaternaris}(0,\beta;\gamma,\beta-\gamma-\delta).\EN

Because the type of any additive code $\codi$ is uniquely defined
by the type of its dual code, for the sake of brevity we further
will use the notation an {\it additive code $\codi$ of dual type
($\alpha,\beta;\gamma',\delta'$)} which means that its additive
dual code ${\cal C}^\perp$ is an additive code of type
($\alpha,\beta;\gamma',\delta'$).

\subsection{Additive extended perfect codes}

In all this paper we will take the permission to call
\textit{additive perfect codes} to additive codes such that, after
the Gray map, give perfect $\add$-linear codes. Also, we will call
{\it additive extended perfect codes} to additive codes such that,
after the Gray map, give a code with the parameters of an extended
perfect $\add$-linear code.

Apart from the linear binary case, so the case when $\beta=0$,
there are two different kinds of additive extended perfect codes,
those with $\alpha \neq 0$ and those with $\alpha=0$. We will
distinguish between these two cases because the constructions are
different.

\begin{theo} \cite{BR98} \label{ExistPerfectZ2Z4LinearCodes} For
each natural number $r$, such that $2\leq r\leq t\leq 2r$, there
exists a unique (up to isomorphism) perfect $\add$-linear code $C$
of binary length $n=2^t-1\geq 15$, such that the $\add$-dual code
of $C$ is of type $(\alpha,\beta;\gamma,\delta)$, where $\alpha
=2^r-1, \beta =2^{t-1}-2^{r-1}, \gamma=2r-t$ and $\delta=t-r$.
\end{theo}

After this theorem we can write the following table (see
\cite{BR98}):

\begin{footnotesize}
\begin{center}
\begin{tabular}{|c|l|l|}
\hline
$t$ & $r$ & ($\alpha,\beta$)  \\
\hline
2    & 2  &  $(3,0)$           \\
3    & 2,3  & $(3,2)$, $(7,0)$           \\
4    & 2,3,4& $(3,6)$, $(7,4)$, $(15,0)$  \\
5    & 3,4,5& $(7,12)$, $(15,8)$, $(31,0)$\\
6    & 3,4,5,6& $(7,28)$, $(15,24)$, $(31,16)$,
$(63,0)$ \\
...  & ...   & ... \\
\hline
\end{tabular}
\end{center}
\end{footnotesize}

Note that for length 7 the two codes are isomorphic and for length
greater than 7 all the codes in the table are non-isomorphic and
unique. The number of non-isomorphic perfect $\add$-linear codes
of length $n=2^t-1$ is $\dd \left\lfloor
\frac{t+2}{2}\right\rfloor$ for all $t>3$, and it is 1 for $t=2$
and $t=3$.

For any $r$ and $t\geq 4$ such that $2\leq r\leq t \leq 2r$, there
is exactly one extended perfect $\add$-linear code $C'$ with
$\alpha=2^r$ and $\beta=2^{t-1}-2^{r-1}$, up to isomorphism. The
corresponding additive codes of these perfect (extended perfect)
$\add$-linear codes are additive perfect (extended perfect) codes
with $\alpha\neq 0$.

Notice that for these extended perfect $\add$-linear codes (or the
corresponding additive extended perfect codes), we do not need to
specify the parameter $\kappa$ because $\kappa=\gamma$ (see
\cite{BF06}), so we just talk about  additive perfect codes of
type ($\alpha,\beta;\gamma,\delta$). Also notice that in this case
given an  additive perfect code $\codi$ of type
($\alpha,\beta;\gamma,\delta$), the additive dual code ${\cal
C}^\perp$ is of type ($\alpha,\beta;\alpha-\gamma,\beta-\delta$).

\begin{ex} 
There are three non-isomorphic perfect $\add$-linear codes of
length $15$. They exist for $(r=2,t=4), (r=3,t=4)$ and
$(r=4,t=4)$. For the case $(r=3,t=4)$ we have the perfect
$\add$-linear code $C$ of dual type $(7,4;2,1)$ with the following
parity-check matrix: $$ \left (\begin{array}{rrrrrrr|rrrr}
0&0&0&1&1&1&1&0&0&2&2\\
0&1&1&0&0&1&1&0&2&0&2\\\hline 1&0&1&0&1&0&1&1&1&1&1
\end{array}\right ). $$ The extended perfect $\add$-linear code
$C'$ for this case is of dual type $(8,4;3,1)$ with the following
parity-check matrix: $$ \left (\begin{array}{rrrrrrrr|rrrr}
1&1&1&1&1&1&1&1&2&2&2&2\\
0&0&0&0&1&1&1&1&0&0&2&2\\
0&0&1&1&0&0&1&1&0&2&0&2\\\hline 0&1&0&1&0&1&0&1&1&1&1&1
\end{array}\right ). $$\end{ex}

\begin{theo} \cite{Kro01} \label{ExistPerfectZ4LinearCodes} For
each $\delta\in\{1,\ldots,\lfloor (t+1)/2 \rfloor\}$ there exists
a unique (up to isomorphism) extended perfect $\Z_4$-linear code
$C'$ of binary length $n+1=2^t\geq 16$, such that the $\Z_4$-dual
code of $C'$ is of type $(0,\beta;\gamma,\delta)$, where $\beta
=2^{t-1}$ and $\gamma=t+1-2\delta$. \end{theo}

\begin{ex} In the case of length $n+1=32$, there are three
non-isomorphic extended perfect $\Z_4$-linear codes, since we have
three possible parameters: $\delta=1$, $\delta=2$ and $\delta=3$.
The following matrix is the parity-check matrix of the code $C'$
for $\delta=2$ (also notice that $\beta=16$ and $\gamma=2$):

$$\left (\begin{array}{cccccccccccccccc}
0&0&0&0&0&0&0&0&2&2&2&2&2&2&2&2 \\
0&0&0&0&2&2&2&2&0&0&0&0&2&2&2&2 \\\hline
1&1&1&1&1&1&1&1&1&1&1&1&1&1&1&1 \\
0&1&2&3&0&1&2&3&0&1&2&3&0&1&2&3 \end{array} \right ).$$ \end{ex}

The corresponding additive codes of the extended perfect
$\Z_4$-linear codes are additive extended perfect codes with
$\alpha=0$. It was established in \cite{BF03} that if the code
$C'$ is an extended perfect $\Z_4$-linear code of binary length
$n+1=2^t\geq 16$, then the punctured code is not a perfect
$\add$-linear code, up to the extended Hamming code of length 16.

The study of additive extended perfect codes is absolutely
different if they come from the extended code of a perfect
$\add$-linear code as in Theorem~\ref{ExistPerfectZ2Z4LinearCodes}
or from extended perfect $\Z_4$-linear codes as in
Theorem~\ref{ExistPerfectZ4LinearCodes}. Note that, in the first
case, the vector with binary ones in the binary part and
quaternary twos in the quaternary part is always in both codes,
the additive extended perfect code and its additive dual code.
However, in the second case, the quaternary all-ones vector is
always in these two codes, the additive extended perfect code and
its additive dual code.

\section{Intersection of additive codes}\label{IntersectionQuaternaryCodes}

In this section we consider the intersection problem for generic
additive codes, with the same length and the same parameters
$\alpha$ and $\beta$.

 We will use the same starting point as in \cite{EV98} and \cite{BYE97}.
Let $\codi_1, \codi_2$ be two additive codes. From~the well-known
second theorem of isomorphism for groups, we can write: \EQ
\label{eqdual}\langle \codi_1, \codi_2 \rangle \big / \codi_1
\cong \codi_2 \big/ (\codi_1\cap \codi_2).\EN

\begin{lemm}\label{second} Let $\codi_1, \codi_2$ be two additive
codes, then $\langle \codi_1^\perp , \codi_2^\perp \rangle =
(\codi_1 \cap \codi_2)^\perp$. \end{lemm} \begin{demo} It is
straightforward to see that $\langle \codi_1^\perp, \codi_2^\perp
\rangle \subset
 (\codi_1 \cap
\codi_2)^\perp$. Moreover, the code $\codi_1 \cap \codi_2$ is the
largest subgroup of both codes $\codi_1$ and $\codi_2$ and, hence,
the code $(\codi_1 \cap \codi_2)^\perp$ is the lowest group
containing both codes $\codi_1^\perp$ and $\codi_2^\perp$. Any
other group containing both codes $\codi_1^\perp$ and
$\codi_2^\perp$ also contains $(\codi_1 \cap \codi_2)^\perp$. This
is the case for the group  $\langle \codi_1^\perp , \codi_2^\perp
\rangle$. \end{demo}

Like in the binary case (see~\cite{EV98}), the above lemma allows
us the following interpretation. Let ${\cal H}_1$ and ${\cal H}_2$
be parity-check matrices of $\codi_1$ and
$\codi_2$, respectively. Then, $\left(\begin{array}{c}{\cal H}_1\\
{\cal H}_2\end{array}\right)$ is a parity-check matrix for the
intersection code $\codi_1 \cap\codi_2$. For the sake of brevity
as in \cite{EV98} we further denote this last matrix by ${\cal
H}_1\parallel {\cal H}_2$.


We begin considering the case $\alpha=0$, and after that, we
investigate the additive case, which is more complicated.

\begin{prop}\label{lemmaBoundsQuaternaryGeneral} For any two
quaternary linear codes \,$\codi_1$ \, and \, $\codi_2$ \, of type
($0,\beta;\gamma_1,\delta_1$) and ($0,\beta;\gamma_2,\delta_2$)
respectively, the  code $\langle \codi_1,\codi_2\rangle$ is  a
quaternary linear code of type ($0,\beta;\gamma,\delta$), where
\begin{equation}\label{BoundsQuaternaryDelta}
 \delta \in
 \{max(\delta_1,\delta_2),\ldots,min(\delta_1+\delta_2,\beta)\} \quad \textrm{and}
\end{equation} \begin{equation}\label{BoundsQuaternaryGamma}
max\big( \delta, max(\gamma_1+\delta_1, \gamma_2+\delta_2)\big)
\leq \gamma  +\delta \leq min(\gamma_1+
\gamma_2+\delta_1+\delta_2,\beta). \end{equation}
 \end{prop}

\begin{demo} Let ${\cal G}_1$ and  ${\cal G}_2$ be  generator
matrices of arbitrary quaternary linear  codes ${\cal C}_1$ and
${\cal C}_2$ of length $\beta$ and types
$(0,\beta;\gamma_1,\delta_1)$ and $(0,\beta;\gamma_2,\delta_2)$.
Consider the  matrix ${\cal G}_1\parallel {\cal G}_2$, which is a
generator matrix of the code $\langle \codi_1,\codi_2\rangle$ of
type ($0,\beta;\gamma,\delta$). After some additive
transformations we get from the matrix ${\cal G}_1\parallel {\cal
G}_2$ a quaternary matrix, denoted by ${\cal G}$, with $\gamma$
rows of order two and $\delta$ rows of order four. It is not
difficult to see that $$max(\delta_1,\delta_2)\leq \delta.$$

Each codeword of order two in the matrix ${\cal G}$ is a linear
combination of rows of order two in the matrices ${\cal G}_1 $ and
${\cal G}_2$, so the total number $\gamma + \delta$ of rows of
order two in the matrix ${\cal G}$ is not more than $\gamma_1
+\delta_1 + \gamma_2 + \delta_2$, i.e. $\gamma + \delta\leq
\gamma_1 +\delta_1 + \gamma_2 + \delta_2$. Moreover,
$\gamma+\delta$ is not greater than the number $\beta$ of coordinates.


On the other hand, $\delta\leq \delta_1  + \delta_2$ because in
some cases there are rows of order two of the matrix ${\cal G}$
that can be obtained by additive combinations of rows of order
four in the matrix ${\cal G}_1\parallel {\cal G}_2$. Also, it must
be $\delta\leq \beta$.

Let us fix $s$ from the set
$\{0,1,\ldots,min(\delta_1,\delta_2)\}$ and suppose that $s$ rows
of order four among all $\delta_1  + \delta_2$ rows of order four
in the matrix ${\cal G}_1\parallel {\cal G}_2$ can be obtained by
linear combinations of rows of the matrix ${\cal G}$. Then, we
immediately get $\delta = \delta_1+\delta_2-s$  independent rows
of order four in the matrix ${\cal G}$. Taking into account this,
we get not less than $max(\gamma_1+\delta_1,
\gamma_2+\delta_2)-\delta$ rows of order two in the matrix ${\cal
G}$. Using the fact that in some cases the number
$max(\gamma_1+\delta_1, \gamma_2+\delta_2)-\delta$ can be  less
than zero, we get the lower bound in
(\ref{BoundsQuaternaryGamma}). \end{demo}

Now we can generalize this last proposition to cover all the
additive codes.

\begin{prop}\label{lemmaBoundsAdditiveGeneral} For any two
additive codes $\codi_1$ and $\codi_2$ of type
($\alpha,\beta;\gamma_1,\delta_1;\kappa_1$) and
($\alpha,\beta;\gamma_2,\delta_2;\kappa_2$) respectively, the code
$\langle \codi_1,\codi_2\rangle$ is  an additive code of type
($\alpha,\beta;\gamma,\delta;\kappa$), where
\begin{equation}\label{nou}
 \delta \in
 \{max(\delta_1,\delta_2),\ldots,min(\delta_1+\delta_2,\beta)\},
\end{equation} \begin{equation}\label{deu} max\big( \delta,
max(\kappa_1+\delta_1, \kappa_2+\delta_2)\big) \leq \kappa+\delta
\leq min(\kappa_1+ \kappa_2+\delta_1+\delta_2,\alpha+\beta)
\end{equation} \begin{equation}\label{onze}  \textrm{and} \quad \kappa+\delta
\leq \gamma +\delta \leq min(\gamma_1+
\gamma_2+\delta_1+\delta_2,\alpha+\beta). \end{equation}
 \end{prop}

\begin{demo} We use the same argumentation than in
Proposition~\ref{lemmaBoundsQuaternaryGeneral}. Taking into
account that always $\kappa\leq \gamma$ we can easily get the
equations (\ref{nou}) and (\ref{onze}).

Now, to obtain (\ref{deu}) we construct two auxiliaries codes
$\codi_1'$ and $\codi_2'$ from the given ones. In the generator
matrices of the codes $\codi_1$ and $\codi_2$ erase, respectively,
the $\gamma_1-\kappa_1$ and $\gamma_2-\kappa_2$ rows which are
dependent on the rest of rows when we restrict them to the binary
part. These new codes $\codi_1'$ and $\codi_2'$ have the same
lengths as the codes $\codi_1$ and $\codi_2$ and are of type
($\alpha,\beta;\kappa_1,\delta_1;\kappa_1$) and
($\alpha,\beta;\kappa_2,\delta_2;\kappa_2$), respectively.

We have $\langle \codi_1',\codi_2'\rangle \subseteq \langle
\codi_1,\codi_2\rangle$ and using the same argumentation than in
Proposition~\ref{lemmaBoundsQuaternaryGeneral} we get the
equation~(\ref{deu}), since the parameter $\gamma$ in $\langle
\codi_1',\codi_2'\rangle$ coincides with the parameter $\kappa$ in
$\langle \codi_1,\codi_2\rangle$. \end{demo}

We finish this section describing the intersection numbers for
additive codes. This means to describe the intersection code just
talking about the cardinality and not about the abelian group
structure. Next lemma allows us to compute the size for the
additive dual code.

\begin{lemm}\label{lemmaDualQuaternary} For any two additive codes
$\codi_1$ and $\codi_2$ of  type
($\alpha,\beta;\gamma_1,\delta_1;\kappa_1$) and
($\alpha,\beta;\gamma_2,\delta_2;\kappa_2$) respectively, the code
$\langle\codi_1,\codi_2\rangle$ is  an additive code of type
($\alpha,\beta;\gamma,\delta;\kappa$), where the size
$|\langle\codi_1,\codi_2\rangle|=2^{\gamma}4^{\delta}$ satisfies
the conditions:
\begin{equation}\label{lemmaDualQuaternaryInequalities}
max(\kappa_1+\delta_1, \kappa_2+\delta_2)+max(\delta_1,\delta_2)
\leq \gamma+2\delta \leq \mu, \end{equation} where $\mu =
min(\gamma_1+\gamma_2+2(\delta_1+\delta_2),
\gamma_1+\gamma_2+\delta_1+\delta_2+\beta,\delta_1+\delta_2+\alpha+\beta,\alpha+2\beta)$.
\end{lemm}

\begin{demo} By Proposition \ref{lemmaBoundsAdditiveGeneral} we
have $$\gamma +\delta \leq min(\gamma_1+
\gamma_2+\delta_1+\delta_2,\alpha+\beta) \quad \textrm{and}$$
$$\delta \leq min(\delta_1+\delta_2,\beta).$$ Therefore $$\gamma
+2\delta \leq \mu,$$ where $\mu =
min(\gamma_1+\gamma_2+2(\delta_1+\delta_2),
\gamma_1+\gamma_2+\delta_1+\delta_2+\beta,\delta_1+\delta_2+\alpha+\beta,\alpha+2\beta)$.

To get the lower bound for $\gamma +2\delta$ we consider the lower
bound in (\ref{deu}) and distinguish two cases.

\noindent Case 1: Let $max(\kappa_1+\delta_1,
\kappa_2+\delta_2)\leq \delta$.  From~(\ref{deu}) we get $0\leq
\kappa\leq \gamma$. Last two inequalities and the lower bound
$max(\delta_1,\delta_2)\leq \delta$ give us
$$max(\kappa_1+\delta_1, \kappa_2+\delta_2)+max(\delta_1,\delta_2)
\leq \gamma+2\delta.$$

\noindent Case 2: For the case $max(\kappa_1+\delta_1,
\kappa_2+\delta_2)\geq \delta$ we have from (\ref{deu}) and
(\ref{onze}) the following inequalities: $max(\kappa_1+\delta_1,
\kappa_2+\delta_2) \leq \kappa + \delta \leq \gamma+\delta$ and,
using the bound $max(\delta_1,\delta_2)\leq \delta$,
 we immediately get the lower
bound in (\ref{lemmaDualQuaternaryInequalities}).
\end{demo}

\begin{theo}\label{intercestionNumberQuaternary} For any two
additive codes $\codi_1$ and $\codi_2$ of dual  type
($\alpha,\beta;\gamma_1,\delta_1;\kappa_1$) and
($\alpha,\beta;\gamma_2,\delta_2;\kappa_2$) respectively, it is
true that $$ 2^{\alpha+2\beta-\mu} \leq \eta(\codi_1, \codi_2)
\leq 2^{\alpha+2\beta -max(\kappa_1+\delta_1,
\kappa_2+\delta_2)-max(\delta_1,\delta_2)}, $$ where $\mu =
min(\gamma_1+\gamma_2+2(\delta_1+\delta_2),
\gamma_1+\gamma_2+\delta_1+\delta_2+\beta,\delta_1+\delta_2+\alpha+\beta,\alpha+2\beta)$.
\end{theo} \begin{demo} Using Lemma \ref{lemmaDualQuaternary} one
can easily get
\begin{equation}\label{intercestionNumberQuaternaryInequalities}
2^{max(\kappa_1+\delta_1,
\kappa_2+\delta_2)+max(\delta_1,\delta_2)} \leq
|\langle\codi_1^\perp, \codi_2^\perp\rangle| \leq 2^{\mu}.
\end{equation}
By Lemma \ref{second} and equation (\ref{McWilltransform}) we have
$$|\langle\codi_1^\perp, \codi_2^\perp\rangle| = |(\codi_1 \cap
\codi_2)^\perp| = 2^{\alpha+2\beta}/|(\codi_1 \cap \codi_2)|.$$
Hence, $\eta(\codi_1, \codi_2) =
2^{\alpha+2\beta}/|\langle\codi_1^\perp, \codi_2^\perp\rangle|$
and the statement follows. \end{demo}

\section{Intersection of  quaternary linear perfect\\
codes}\label{qlpc}

In this section we consider additive extended perfect codes such
that $\alpha=0$, which will also be called {\it quaternary linear
perfect codes}. We investigate the intersection problem for such
codes and the abelian group structure for their intersection
codes. In Subsection~\ref{number} we consider the intersection
problem and in Subsection~\ref{structure} the abelian group
structure of these intersection codes.

\subsection{Intersection problem of quaternary linear perfect codes}\label{number}

All statements presented in the previous section are valid for the
quaternary linear perfect codes with some modifications. We are
going to omit some proofs indicating some specific properties of
these codes. So, taking into account that the quaternary all-ones
vector always belongs to the quaternary dual of any quaternary
linear perfect code, we immediately get from Proposition
\ref{lemmaBoundsQuaternaryGeneral}:

\begin{prop}\label{lemmaBoundsQuaternary} For any two  quaternary
linear perfect codes such that their quaternary dual codes are
$\codi_1$ and $\codi_2$ of type ($0,\beta;\gamma_1,\delta_1$) and
($0,\beta;\gamma_2,\delta_2$) respectively, with $\beta=2^{t-1}$
and $t\geq 4$, the  code $ \langle\codi_1,\codi_2\rangle$ is  a
quaternary linear code of type ($0,\beta;\gamma,\delta$), where
\begin{equation}
 \delta \in \{max(\delta_1,\delta_2),\ldots, \delta_1+\delta_2-1 \}
\quad \textrm{and} \end{equation} \begin{equation} max\big(
\delta, max(\gamma_1+\delta_1, \gamma_2+\delta_2)\big) \leq \gamma
+\delta \leq  \gamma_1+ \gamma_2+\delta_1+\delta_2-1.
\end{equation} \end{prop} \begin{demo} The quaternary linear
perfect codes satisfy (see
Theorem~\ref{ExistPerfectZ4LinearCodes}) $\beta=2^{t-1}$,
$\gamma_1+2\delta_1=\gamma_2+2\delta_2=t+1$ and
$\delta_1,\delta_2\in \{1,\ldots,\lfloor(t+1)/2\rfloor\}$.

The quaternary all ones vector is always in the quaternary dual,
which means that the upper bound for $\delta$ is not
$min(\delta_1+\delta_2,\beta)$ but
$min(\delta_1+\delta_2-1,\beta)$. Since $t\geq 4$ using Theorem
\ref{ExistPerfectZ4LinearCodes} we have $\delta_1+\delta_2-1 \leq
t\leq \beta$, so  we can write $\delta_1+\delta_2-1$ as the upper
bound for $\delta$ in equation (\ref{BoundsQuaternaryDelta}).

Moreover taking into account that
$\gamma_1+2\delta_1=\gamma_2+2\delta_2=t+1$ and $\delta_1 \geq 1,
\delta_2 \geq 1$ we get $\gamma_1+\gamma_2+\delta_1+\delta_2-1 =
2(t+1)-\delta_1-\delta_2-1\leq 2t-1$. Again, since $t\geq 4$ we
have $2t-1<\beta$, so we can write
$\gamma_1+\gamma_2+\delta_1+\delta_2-1$ as the upper bound for
$\gamma+\delta$ in equation (\ref{BoundsQuaternaryGamma}). This
proves the statement. \end{demo}

Next two theorems give  us the solution of the intersection
problem for quaternary linear perfect codes. First, we present the
lower and the upper bounds. Then, we  show that there exist such
codes for any possible intersection number between these bounds.

\begin{theo}\label{boundsQuaternary} For any $t\geq 3$ and any
quaternary linear perfect codes $\codi_1$ and $\codi_2$ of length
$\beta=2^{t-1}$, it is true  that $$2^{2\beta-2t} \leq
\eta(\codi_1,\codi_2) \leq 2^{2\beta-t-1}.$$ \end{theo}

\begin{demo} We know that for the quaternary linear perfect codes
$\codi_1$ and $\codi_2$ the quaternary all-ones vector is in their
quaternary dual codes, so the statement in
Theorem~\ref{intercestionNumberQuaternary} is $$2^{2\beta-\mu}
\leq \eta(\codi_1, \codi_2) \leq 2^{2\beta -max(\gamma_1+\delta_1,
\gamma_2+\delta_2)-max(\delta_1,\delta_2)}, $$ where using
Proposition \ref{lemmaBoundsQuaternary} we immediately get $\mu
=2t.$
Also, by Theorem~\ref{ExistPerfectZ4LinearCodes}, we know that
$\beta=2^{t-1}$ and $\gamma_1+2\delta_1=\gamma_2+2\delta_2=t+1$,
so for $t\geq 3$ we have:
$$max(\gamma_1+\delta_1,\gamma_2+\delta_2)+max(\delta_1,\delta_2)=t+1+|\delta_2-\delta_1|.$$
Therefore
$$2^{2\beta -max(\gamma_1+\delta_1,
\gamma_2+\delta_2)-max(\delta_1,\delta_2)}=2^{2\beta-(t+1)-|\delta_2-\delta_1|}\leq
2^{2\beta-t-1},$$ so the statement follows. \end{demo}

The lower bound given by Theorem \ref{boundsQuaternary} is an even
power of two. So, comparing with the intersection problem for
binary Hamming codes, see (\ref{EtzionVardyHamIntersection})
above, it is impossible to get two extended perfect $\Z_4$-linear
codes of length $2\beta=2^t$ ($t\geq 3$) with intersection number
$2^{2\beta-2t-1}$.

\begin{theo}\label{teorema5} For any $t\geq 3$ there exist two
quaternary linear perfect codes $\codi_1$ and $\codi_2$ of  length
$\beta=2^{t-1}$, such that $\eta(\codi_1, \codi_2) =
2^{2\beta-l}$, where $l$ is any value from $t+1$ to $2t$.
\end{theo}

\begin{demo} By Theorem \ref{boundsQuaternary}, the minimum and
maximum intersection  numbers for  quaternary linear perfect codes
$\codi_1$ and $\codi_2$ of length $\beta=2^{t-1}$ are
$2^{2\beta-2t}$ and $2^{2\beta-t-1}$, respectively.

For $t=3$, we have to find quaternary linear perfect codes of
length $\beta=4$ with intersection numbers 16, 8 and 4. Let
$\codi_1$ and $\codi_2$ be the quaternary linear perfect codes
with the parity-check matrices
$$\left (\begin{array}{cccc} \hline
1&1&1&1 \\
0&1&2&3 \end{array} \right ) \quad \textrm{and} \quad \left
(\begin{array}{cccc}
0&0&2&2\\
0&2&0&2\\ \hline 1&1&1&1\end{array} \right ),$$ respectively.
Then, it is easy to see that $\eta(\codi_1,\codi_1)=16$,
$\eta(\codi_1,\codi_2)=8$ and $\eta(\codi_1,\pi(\codi_1))=4$,
where $\pi=(1,2)$.

For $t\geq 4$, we consider the parity-check matrix of the
quaternary linear perfect code of length $\beta$ with $\gamma =
t-1$ and $\delta = 1$. This parity-check matrix can be represented
as the quaternary matrix $${\cal H}=\left(\begin{array}{c}2H
\\\hline 1\dots 1\end{array}\right),$$ where $H$ is the matrix
with the first column the all-zeroes vector of length $t-1$  and
the rest of the columns from the parity-check matrix of a binary
Hamming code of length $\beta -1$. By the classification of
intersection numbers of binary Hamming codes given by Etzion and
Vardy in \cite{EV98}, the rank of the matrix $H\parallel \pi (H)$
can vary from $t-1$ till $2(t-1)$ for different permutations $\pi$
of length $\beta$. Then, there exist quaternary linear codes
$\codi_1$ and $\codi_2$ of length $\beta$ with parity-check
matrices ${\cal H}$ and $\pi({\cal H})$ respectively, such that
the code $\langle \codi_1^\perp, \codi_2^\perp \rangle$ with
generator matrix ${\cal H} \parallel \pi ({\cal H})$ is of type
$(0,\beta;r,1)$ for all $ r\in \{t-1,\ldots, 2t-2\}$. Since
$|\codi_1 \cap \codi_2|\cdot |\langle \codi_1^\perp, \codi_2^\perp
\rangle|=2^{2\beta}$ and $|\langle \codi_1^\perp, \codi_2^\perp
\rangle|=2^r4$, we have the intersection numbers $\eta(\codi_1,
\codi_2) =2^{2\beta-(r+2)} =2^{2\beta-l}$, where $l=r+2$ is any
value from $t+1$ to $2t$. \end{demo}

\subsection{The abelian group structure for the intersection\\ of
quaternary linear perfect codes}\label{structure}

To investigate the abelian group structure for the intersection of
quaternary linear perfect codes it will be helpful  the following
statement, which we can get immediately from Proposition
\ref{lemmaBoundsQuaternary}.

\begin{theo} \label{intersectionStructureBounds} For any two
quaternary linear perfect codes $\codi_1$ and $\codi_2$ of dual
 type ($0,\beta;\gamma_1,\delta_1$) and
($0,\beta;\gamma_2,\delta_2$) respectively, with $\beta=2^{t-1}$
and $t\geq 4$, the intersection code $\codi_1 \cap \codi_2$ is of
dual type ($0,\beta;\gamma,\delta$), where $\gamma$ and $\delta$
satisfy the bounds given by
Proposition~\ref{lemmaBoundsQuaternary}. \end{theo}

From~this theorem, it is easy to compute the type of the
intersection code $\codi_1 \cap \codi_2$ using the fact that if a
quaternary linear code has parameters ($0,\beta;\gamma,\delta$),
then its quaternary dual code has parameters
($0,\beta;\gamma,\beta - \gamma - \delta$).

\bigskip Next we will show that there exist quaternary linear
perfect codes of length $\beta=2^{t-1}$ for any $t\geq 4$, with
intersections of type ($0,\beta;\gamma,\delta$) for all possible
$\gamma$ and $\delta$ between the bounds given by Theorem
\ref{intersectionStructureBounds}. In Example \ref{ExBeta4}, for
quaternary linear perfect codes of length $\beta=4$ ($t=3$), we
show which intersections codes with parameters between these
bounds are possible.

\begin{ex} \label{ExBeta4} For $\beta=4$ ($t=3$) there are two
isomorphic quaternary linear perfect  codes $\codi_1$ and
$\codi_2$ given by $\delta=1$ and $\delta=2$, so of dual types
($0,4;2,1$) and ($0,4;0,2$), respectively. We can take
$${\cal H}_1=\left(\begin{array}{cccc} 0&0&2&2\\
0&2&0&2 \\ \hline 1&1&1&1\end{array}\right) \quad \textrm{and}
\quad {\cal H}_2=\left(\begin{array}{cccc} \hline 1&1&1&1 \\
0&1&2&3\end{array}\right)$$ as parity-check matrices of $\codi_1$
and $\codi_2$, respectively.

By an exhaustive search, the intersection code $\codi_1 \cap
\pi(\codi_1)$ with parity-check matrix ${\cal H}_1 \parallel
\pi({\cal H}_1)$ is of dual type ($0,4;2,1$) for any permutation
$\pi$. On the other hand, the intersection code $\codi_2 \cap
\pi(\codi_2)$ is either of dual type ($0,4;0,2$) or ($0,4;0,3$).
For example, taking $\pi=Id$ and $\pi=(1,2)$, we find that
$$\left(\begin{array}{cccc} \hline 1&1&1&1 \\
0&1&2&3\end{array}\right) \quad \textrm{and} \quad
\left(\begin{array}{cccc} \hline 1&1&1&1 \\
0&1&2&3 \\ 1&0&2&3 \end{array} \right)$$ are parity-check matrices
of these two intersection codes. Finally, the intersection code
$\codi_1 \cap \pi(\codi_2)$ is always of dual type ($0,4;1,2$) for
any permutation $\pi$.\end{ex}

\begin{lemm} \label{lemmaInterStructure2beta} Let $\codi_1$ and
$\codi_2$ be quaternary linear perfect codes of dual type
$(0,\beta;\gamma_1,\delta_1)$ and $(0,\beta;\gamma_2,\delta_2)$
respectively, such that $\codi_1 \cap \codi_2$ is of dual type
$(0,\beta; i,j)$, $\beta \geq 4.$ Then, there exist two quaternary
linear perfect codes  of dual type $(0,2\beta;
\gamma_1+1,\delta_1)$ and $(0,2\beta; \gamma_2+1,\delta_2)$ with
intersection codes of dual type $(0,2\beta; i',j)$
for $i'\in \{i+1,i+2\}$. \end{lemm}

\begin{demo} Let ${\cal H}_1$ and ${\cal H}_2$ be parity-check
matrices of the quaternary linear perfect codes $\codi_1$ and
$\codi_2$, respectively. The matrices $$\left (\begin{array}{cc}
0\ldots 0 & 2\ldots 2 \\
{\cal H}_1& {\cal H}_1 \end{array} \right ) \quad \textrm{and}
\quad \left (\begin{array}{cc}
0\ldots 0 & 2\ldots 2 \\
{\cal H}_2& {\cal H}_2 \end{array} \right )$$ are parity-check
matrices of quaternary linear perfect codes ${\cal D}_1$ and
${\cal D}_2$ of dual type ($0,2\beta;\gamma_1+1,\delta_1$) and
($0,2\beta;\gamma_2+1,\delta_2$), respectively. Since the
intersection code $\codi_1 \cap \codi_2$ is of dual type
($0,\beta;i,j$), the intersection code ${\cal D}_1 \cap {\cal
D}_2$ is of dual type ($0,2\beta; i+1,j$). Moreover, taking the
permutation $\pi=(1,\beta+1)$, the intersection code ${\cal D}_1
\cap \pi({\cal D}_2)$ is of dual type ($0,2\beta;i+2,j$), because
we are adding the row $\pi(0,\ldots,0,2,\ldots,2)$ of order two in
the parity-check matrix of this intersection.  \end{demo}

\begin{lemm} \label{lemmaInterStructure4beta} Let $\codi_1$ and
$\codi_2$ be quaternary linear perfect codes of dual type
$(0,\beta;\gamma_1,\delta_1)$ and $(0,\beta;\gamma_2,\delta_2)$
respectively, such that $\codi_1 \cap \codi_2$ is of dual type
$(0,\beta; i,j)$, $\beta \geq 4.$ Then, there exist two quaternary
linear perfect codes  of dual type $(0,4\beta;
\gamma_1,\delta_1+1)$ and $(0,4\beta; \gamma_2,\delta_2+1)$ with
intersection codes of dual type $(0,4\beta; i',j')$
for $ (i',j') \in \{(i,j+1), (i,j+2), (i+1,j+1) \}$.
\end{lemm}

\begin{demo} Let ${\cal H}_1$ and ${\cal H}_2$ be parity-check
matrices of the quaternary linear perfect codes $\codi_1$ and
$\codi_2$, respectively. The matrices $$\left (\begin{array}{cccc}
{\cal H}_1& {\cal H}_1 & {\cal H}_1 & {\cal H}_1 \\
0\ldots 0 & 1\ldots 1  & 2\ldots 2 & 3\ldots 3 \\
\end{array} \right ) \ \textrm{and} \  \left (\begin{array}{cccc}
{\cal H}_2& {\cal H}_2 & {\cal H}_2 & {\cal H}_2 \\
0\ldots 0 & 1\ldots 1  & 2\ldots 2 & 3\ldots 3 \\
\end{array} \right )$$ are parity-check matrices of quaternary
linear perfect codes ${\cal D}_1$ and ${\cal D}_2$ of dual type
($0,4\beta;\gamma_1,\delta_1+1$) and
($0,4\beta;\gamma_2,\delta_2+1$), respectively. Since the code
$\codi_1 \cap \codi_2$ is of dual type ($0,\beta;i,j$), the
intersection code ${\cal D}_1 \cap {\cal D}_2$ is of dual type
($0,4\beta; i,j+1$). Taking for example the permutation
$\pi=(1,\beta+1)$, the intersection code ${\cal D}_1 \cap
\pi({\cal D}_2)$ is of dual type ($0,4\beta;i,j+2$), because we
are adding two new independent rows of order four, $v=(0,\ldots,
0, 1,\ldots, 1, 2,\ldots, 2, 3,\ldots,3)$ and $\pi(v)$,
 to the parity-check matrix of the
intersection. Finally, taking the permutation
$\sigma=(1,2\beta+1)$, the intersection code ${\cal D}_1 \cap
\sigma({\cal D}_2)$ is of dual type ($0,4\beta;i+1,j+1$), because
the rows, $v$ and $\sigma(v)$, in the parity-check matrix are
equivalent to the rows $v$ and $(2,0,\ldots,0,2,0,\ldots,0)$ of
order four and two, respectively. \end{demo}

\begin{lemm} \label{lemmaInterStructureBeta} For all $m\geq 2$
there exist two quaternary linear perfect codes $\codi_1$ and
$\codi_2$ of dual type ($0,2^{2m};0,m+1$) and ($0,2^{2m};2m,1$)
respectively, such that $\codi_1 \cap \codi_2$ is a quaternary
linear code of dual type $(0,2^{2m}; \gamma, m+1)$, where $\gamma$
is any value from $m$ to $2m$ (so any value given by Theorem
\ref{intersectionStructureBounds}). \end{lemm}

\begin{demo} Let ${\cal H}_1$ and ${\cal H}_2$ be parity-check
matrices of the quaternary linear perfect codes $\codi_1$ and
$\codi_2$ of dual types ($0,2^{2m};0,m+1$) and ($0,2^{2m};2m,1$),
respectively. Assume $${\cal H}_1=\left(\begin{array}{c} \hline
1\dots 1 \\ q_1 \\ \vdots \\ q_m \end{array}\right) \quad
\textrm{and} \quad {\cal H}_2=\left(\begin{array}{c}2H_{2m}
\\\hline 1\dots 1\end{array}\right),$$ where the rows
$q_1,\ldots,q_m$ form a submatrix that has as columns all the
vectors of $\Z_4^m$ ordered lexicographically, and $H_{2m}$ is the
parity-check matrix of an extended binary Hamming code of length
$2^{2m}$ whose columns are also ordered lexicographically.

The code $\codi_1 \cap \pi(\codi_2)$, which has parity-check
matrix ${\cal H}_1 \parallel \pi({\cal H}_2)$, is of dual type
$(0,2^{2m}; \gamma_{\pi},m+1)$, where $\gamma_\pi \in
\{m,\ldots,2m\}$, for any permutation $\pi$ on the set of
coordinates.

The rows $2q_1, \ldots , 2q_{m}$ of order two are included in the
matrix ${\cal H}_2$. So, in the matrix ${\cal H}_1 \parallel {\cal
H}_2$ there are only $m$ independent rows of order two, which
means that $\codi_1 \cap \codi_2$ is a quaternary linear code of
dual type ($0,2^{2m};m,m+1$).

For each $i\in \{1,\ldots,m\}$, there exists a transposition
$\sigma_i$ that fixes all the rows $2q_j$ for $j=1,\ldots,m$,
$j\not =i$, and switches two coordinates that contain different
elements in $2q_i$. Taking the permutation $\pi_i=\sigma_1 \scirc
\sigma_2 \scirc \cdots \scirc \sigma_i$, the intersection code
$\codi_1 \cap \pi_i(\codi_2)$ is of dual type
($0,2^{2m};m+i,m+1$).
\end{demo}

\begin{lemm} \label{lemmaInterStructureBetaGeneral} For all $m\geq
2$ and any $\gamma_1,\delta_1$, such that
$\gamma_1+2\delta_1=2m+2$ and $\delta_1\geq 1$, there exist two
quaternary linear perfect codes $\codi_1$ and $\codi_2$ of dual
type ($0,2^{2m};\gamma_1,\delta_1$) and ($0,2^{2m};0,m+1$)
respectively, such that $\codi_1 \cap \codi_2$ is a quaternary
linear code of dual type $(0,2^{2m}; \gamma , \delta)$, where
$\gamma$ and $\delta$ are any values $$m+1 \leq \delta \leq
\delta_1+m \quad \textrm{and}$$ $$max(0,\gamma_1+\delta_1-\delta)
\leq \gamma \leq \gamma_1+\delta_1-\delta+m$$ (so any values given
by Theorem \ref{intersectionStructureBounds}). \end{lemm}

\begin{demo} By Theorem \ref{ExistPerfectZ4LinearCodes}, we know
that for each $\delta_1 \in \{1,\ldots,m+1\}$ there exists a
non-isomorphic quaternary linear perfect code of dual type
($0,2^{2m};2(m-\delta_1+1),\delta_1$). We will prove the
statement, by induction on $m$ and for each possible $\delta_1$.

First, for $m=2$ we need to show the existence of two perfect
codes of dual type ($0,16;6-2\delta_1,\delta_1$) and ($0,16;0,3$)
with all possible intersections, for each $\delta_1\in \{1,2,3\}$.
For $\delta_1=1$, we have the result by Lemma
\ref{lemmaInterStructureBeta}. For $\delta_1=2$, we use Lemma
\ref{lemmaInterStructure4beta} and the codes constructed in
Example \ref{ExBeta4} of dual types ($0,4;2,1$) and ($0,4;0,2$)
with intersection of dual type ($0,4;1,2$). Hence, there exist
codes $\codi_1$ and $\codi_2$ of dual type ($0,16;2,2$) and
($0,16;0,3$) with intersection codes of dual types ($0,16;1,3$),
($0,16;1,4$) and ($0,16;2,3$). Moreover, taking the quaternary
linear perfect codes $\codi_1$ and $\codi_2$ of length 16 with
parity-check matrices $${\cal H}_1=\left
(\begin{array}{cccccccccccccccc}
0&0&0&0&0&0&0&0&2&2&2&2&2&2&2&2 \\
0&0&0&0&2&2&2&2&0&0&0&0&2&2&2&2 \\\hline
1&1&1&1&1&1&1&1&1&1&1&1&1&1&1&1 \\
0&1&2&3&0&1&2&3&0&1&2&3&0&1&2&3 \end{array} \right )$$ and ${\cal
H}_2$ below, respectively, and the permutations
$\pi_1=(2,4)(3,5)$, $\pi_2=(1,2)(3,4)(5,8,7,6)$ and
$\pi_3=(1,2)(3,5)$; the intersection codes $\pi_i(\codi_1)\cap
\codi_2$, $i=1,2,3$, are of dual type ($0,16;3,3$), $(0,16;0,4)$
and $(0,16;2,4)$, respectively. This gives all possible values for
$m=2$ and $\delta_1=2$. Finally, we prove the result for $m=2$ and
$\delta_1=3$. Again by Example \ref{ExBeta4}, there exist
(isomorphic) perfect codes of dual type ($0,4;0,2$) with
intersection codes of dual type ($0,4;0,2$) and ($0,4;0,3$). Then,
by Lemma \ref{lemmaInterStructure4beta}, there exist codes
$\codi_1$ and $\codi_2$ of dual type ($0,16;0,3$) with
intersection codes of dual type ($0,16;i,j$), for all $0\leq i\leq
5-j$ and $3 \leq j \leq 5$, except for the case when $i=2$ and
$j=3$. Moreover, taking the quaternary linear perfect code
$\codi_2$ of length 16 with parity-check matrix $${\cal H}_2=\left
(\begin{array}{cccccccccccccccc} \hline
1&1&1&1&1&1&1&1&1&1&1&1&1&1&1&1 \\
0&0&0&0&1&1&1&1&2&2&2&2&3&3&3&3 \\
0&1&2&3&0&1&2&3&0&1&2&3&0&1&2&3
\end{array} \right )$$ and the permutation
$\pi=(7,13,15)(8,14,16)$, the intersection code $\codi_2\cap
\pi(\codi_2)$ is of dual type ($0,16;2,3$). So, the result is true
for $m=2$.

Now, we assume that the result is true for perfect codes of dual
type ($0,2^{2(m-1)};\gamma_1,\delta_1$) and ($0,2^{2(m-1)};0,m$),
for each $\delta_1 \in \{1,\ldots,m\}$ and
$\gamma_1=2(m-\delta_1)$. Let there exist intersection codes of
dual type ($0,2^{2(m-1)};i',j'$), for all $$m\leq j'\leq
\delta_1+m-1 \quad \textrm{and}$$ $$max(0,\gamma_1+\delta_1-j')
\leq i'\leq \gamma_1+\delta_1-j'+m-1.$$ Then, by Lemma
\ref{lemmaInterStructure4beta}, there exist codes ${\cal D}_1$ and
${\cal D}_2$ of dual type ($0,2^{2m};\gamma_1,\delta_1+1$) and
($0,2^{2m};0,m+1$) with intersection codes of dual type
($0,2^{2m};i,j$), for all $$m+1\leq j\leq (\delta_1+1)+m  \quad
\textrm{and}$$ $$max(0,\gamma_1+(\delta_1+1)-j) \leq i \leq
\gamma_1+(\delta_1+1)-j+m,$$ given any $\delta_1+1\in
\{2,\ldots,m+1\}$. If $\delta_1=1$ then $\gamma_1=2m$ (see Theorem
\ref{ExistPerfectZ4LinearCodes}) and by Lemma
\ref{lemmaInterStructureBeta}, we have the result for two perfect
codes of dual type $(0,2^{2m};\gamma_1,1)$ and $(0,2^{2m};0,m+1)$.
So, the result is true for any $\delta_1\in
\{1,\ldots,m+1\}$.\end{demo}

\begin{theo} \label{theoremgapsquaternary}For all $t\geq 4$ and
any $\gamma_1,\delta_1,\gamma_2,\delta_2$, such that
$\gamma_1+2\delta_1=\gamma_2+2\delta_2=t+1$ and
$\delta_1,\delta_2\geq 1$, there exist two quaternary linear
perfect codes $\codi_1$ and $\codi_2$ of dual type
($0,2^{t-1};\gamma_1,\delta_1$) and
($0,2^{t-1},\gamma_2,\delta_2$) respectively,  such that $\codi_1
\cap \codi_2$ is a quaternary linear code of dual type
$(0,2^{t-1}; \gamma , \delta)$, where
\EQ\label{equation}\begin{split}\delta \in
\{max(\delta_1,\delta_2),\ldots, \delta_1+\delta_2-1 \} \quad
\textrm{and} \\ max\big( \delta, max(\gamma_1+\delta_1,
\gamma_2+\delta_2)\big) \leq \gamma +\delta \leq  \gamma_1+
\gamma_2+\delta_1+\delta_2-1.\end{split}\EN \end{theo}

\begin{demo} By Theorem \ref{ExistPerfectZ4LinearCodes}, we know
that for each $t\geq 4$ there are $\lfloor (t+1)/2 \rfloor$
non-isomorphic quaternary linear perfect codes of length
$\beta=2^{t-1}$. Specifically, for each
$\bdelta\in\{1,\ldots,\lfloor (t+1)/2 \rfloor\}$ there exists one
code of dual type ($0,2^{t-1};t+1-2\bdelta,\bdelta$).

We will prove the statement by induction on $t\geq4$, and we need
to show the result is true for the initial cases $t=4$ and $t=5$.

For $t=4$, we have two non-isomorphic quaternary linear perfect
codes $\codi_1$ and $\codi_2$ given by $\bdelta=1$ and
$\bdelta=2$, respectively, and with
parity-check matrices $${\cal H}_1=\left(\begin{array}{c} 2H_3 \\
\hline
1\ldots1 \\
\end{array} \right) \quad \textrm{and} \quad {\cal H}_2=\left
(\begin{array}{cccccccc} 0&0&0&0&2&2&2&2 \\ \hline
1&1&1&1&1&1&1&1 \\
0&1&2&3&0&1&2&3 \end{array} \right),$$ where $H_3$ is a
parity-check matrix of an extended binary Hamming code of length
8. When $\delta_1=\delta_2=1$, by using the proof of Theorem
\ref{teorema5}, we have intersections of dual type ($0,8;r,1$) for
any $r$ from $\{3,4,5,6\}$. When $\delta_1=\delta_2=2$, using the
quaternary linear perfect code of dual type ($0,4;0,2$), the
intersection codes in Example \ref{ExBeta4} and Lemma
\ref{lemmaInterStructure2beta}, we get all the possible
intersection codes for this case, except the intersections of dual
types ($0,8;3,2$) and ($0,8;0,3$). However, taking the
permutations $\pi=(1,5)(2,4)$ and $\sigma=(1,2)(3,4)$, the codes
$\codi_2 \cap \pi(\codi_2)$ and $\codi_2 \cap \sigma(\codi_2)$ are
of dual types ($0,8;3,2$) and ($0,8;0,3$), respectively. Finally,
when $\delta_1=1$ and $\delta_2=2$, using the perfect codes of
dual types ($0,4;2,1$) and ($0,4;0,2$), the intersection code of
dual type ($0,4;1,2$) in Example \ref{ExBeta4} and Lemma
\ref{lemmaInterStructure2beta}, there exist all possible
intersections codes for this case, except the intersection code of
dual type ($0,8;4,2$). Taking the permutation $\tau=(1,2)(4,5)$,
the code $\tau(\codi_1) \cap \codi_2$ is of this missing type.

Similarly, for $t=5$ we can find, by direct search, all the
possible intersection codes fulfilling the statement. We avoid to
write here the complete list.

Now, we assume the result is true for quaternary linear perfect
codes of length $\beta=2^{t-2}$ and $\beta=2^{t-3}$ ($t>5$). So,
this means we are assuming that for any $\delta_1,\delta_2 \in
\{1,\ldots, \lfloor t/2\rfloor\}$ we have quaternary linear
perfect codes of dual type ($0,2^{t-2};t-2\delta_1,\delta_1$) and
($0,2^{t-2};t-2\delta_2,\delta_2$) with intersection codes of dual
type ($0,2^{t-2};i',j'$), for all $i', j'$ fulfilling
equations~(\ref{equation}).
By Lemma \ref{lemmaInterStructure2beta}, we have quaternary linear
perfect codes of dual type ($0,2^{t-1}; t+1-2\delta_1,\delta_1$)
and ($0,2^{t-1};t+1-2\delta_2,\delta_2$) with intersection codes
of dual type ($0,2^{t-1};i,j$), for all $j=j'$ and $i\in
\{i'+1,i'+2\}$, where $$ max(\delta_1,\delta_2) \leq j \leq
\delta_1+\delta_2-1 \quad \textrm{and}$$
$$max(1,max(t+1-\delta_1,t+1-\delta_2)-j)\leq i \leq
2(t+1)-\delta_1-\delta_2-j-1.$$ So, we obtain all the possible
types for the intersection, except for $i=0$ and $j\geq
max(t+1-\delta_1,t+1-\delta_2)$. But, in this exceptional case, we
have $j\geq max(\delta_1,\delta_2)+1$ because, otherwise, assuming
$j< \delta_1+1$ we would have $\delta_1+1 >t+1-\delta_1$ and
$2\delta_1>t$ which is a contradiction.

Also, from induction hypothesis, we can assume that for any
$\delta_1,\delta_2 \in \{2,\ldots, \lfloor (t+1)/2\rfloor\}$ we
have quaternary linear perfect codes of dual type
($0,2^{t-3};(t-1)-2(\delta_1-1),\delta_1-1$) and
($0,2^{t-3};(t-1)-2(\delta_2-1),\delta_2-1$) with intersection
codes of dual type ($0,2^{t-3};i',j'$), for all $i', j'$
fulfilling equations~(\ref{equation}). By
Lemma~\ref{lemmaInterStructure4beta}, we have quaternary linear
perfect codes of dual type ($0,2^{t-1};t+1-2\delta_1,\delta_1$)
and ($0,2^{t-1};t+1-2\delta_2,\delta_2$) with intersection codes
of dual type ($0,2^{t-1};i,j$), for all $(i,j)\in \{ (i',j'+1),
(i',j'+2), (i'+1,j'+1)\}$, where $$ max(\delta_1,\delta_2)+1 \leq
j \leq \delta_1+\delta_2-1 \quad \textrm{and}$$
$$max(0,max(t+1-\delta_1,t+1-\delta_2)-j)\leq i \leq
2(t+1)-\delta_1-\delta_2-j-1.$$

Notice that when $t$ is even then $\lfloor t/2\rfloor$ coincides
with $\lfloor (t+1)/2\rfloor$ so, in this case the proof is
finished. When $t$ is odd we need to prove the statement for
$\delta_1\in  \{1,\ldots,(t+1)/2\}$ and $\delta_2=(t+1)/2$, which
is straightforward from Lemma
\ref{lemmaInterStructureBetaGeneral}.\end{demo}

\section{Intersection of additive extended perfect codes with
$\alpha~\not=~0$}\label{aepc}

In this section we consider the intersection problem for additive
extended perfect codes such that $\alpha\neq 0$. We also
investigate the abelian group structure for the intersection of
such codes. Again, all statements presented in Section
\ref{IntersectionQuaternaryCodes} are valid for this case with
some small changes.

As we said before, the extended perfect $\Z_4$-linear codes and
the extended codes of the perfect $\add$-linear codes could be
seen as additive extended perfect codes after the Gray map. In the
first case, the quaternary all-ones vector belongs to the code and
also to the quaternary dual code. In the second case, the vector
with binary ones in the binary part and quaternary twos in the
quaternary part is always in the code and also in the additive
dual code.

Given an additive extended perfect code $\codi$ of dual type
($\alpha,\beta;\gamma,\delta$) with $\alpha\neq 0$, we always have
$\alpha+2\beta=2^{t}$; $\gamma+2\delta=t+1$; $\alpha=2^r$ and
$\gamma+\delta =r+1$ (see Theorem
\ref{ExistPerfectZ2Z4LinearCodes}). Hence, given the parameters
$\alpha$ and $\beta$, all the additive extended perfect codes with
these parameters must have the same parameters $\gamma$ and
$\delta$.

\medskip Proposition \ref{lemmaBoundsAdditiveGeneral} is
transformed in the next proposition:

\begin{prop}\label{lemmaBoundsAdditive} For any two additive
extended perfect codes such that their additive dual codes are
$\codi_1$ and $\codi_2$ of type ($\alpha,\beta;\bgamma,\bdelta$)
with $\alpha\not=0$,  the code $\langle\codi_1,\codi_2\rangle$ is
an additive code of type ($\alpha,\beta;\gamma,\delta;\kappa$),
where $$\left \{\begin{array}{lll} \mbox{if } \bdelta =0 & \mbox{
then }& \delta=0 \mbox{ and } \bgamma\leq \kappa=\gamma\leq
2\bgamma
-1,\\
\mbox{if } \bdelta=1 & \mbox{ then } & \delta=1 \mbox{ and
}\bgamma \leq
\kappa \leq \gamma\leq 2\bgamma,\\
\mbox{if }\bdelta >1 & \mbox{ then } & \delta\in
\{\bdelta,\ldots,2\bdelta\} \mbox{ and }\bgamma \leq \kappa \leq
\gamma \leq 2\bgamma+2\bdelta-\delta-1. \end{array}\right.$$
\end{prop}

\begin{demo} If $\codi_1$ and $\codi_2$ are the additive dual
codes of type ($\alpha,\beta;\bgamma,\bdelta$) with
$\alpha\not=0$, then there exist values $2\leq r\leq t\leq 2r$
(see Theorem~\ref{ExistPerfectZ2Z4LinearCodes}) such that
$\alpha=2^r$, $\beta=2^{t-1}-2^{r-1}$, $\bgamma=2r-t+1$ and
$\bdelta=t-r$.

For these codes Proposition~\ref{lemmaBoundsAdditiveGeneral}
becomes \EQ\label{eq1} \delta \in
\{\bdelta,\ldots,min(2\bdelta,\beta)\},\EN \EQ\label{eq2}
max(\delta,\bgamma+\bdelta)\leq \kappa+\delta\leq
min(2(\bgamma+\bdelta)-1,\alpha+\beta)\EN \EQ \label{eq3}
\textrm{and} \quad \kappa+\delta\leq \gamma+\delta\leq
min(2(\bgamma+\bdelta)-1,\alpha+\beta),\EN because the vector with
binary ones in the binary part and quaternary twos in the
quaternary part is always in $\codi_1$ and $\codi_2$.

The lower bound in equation (\ref{eq2}) can be improved. The
$\bgamma$ vectors of order two in $\codi_1$ or $\codi_2$ are
necessarily independent of the $\bdelta$ vectors of order four on
the other code, respectively, so the lower bound becomes
$\bgamma+\delta$. Moreover, for $t\geq3$ we have
$\alpha+\beta=2^{t-1}+2^{r-1}=2^{r-1}(2^{t-r}+1)\geq 2r+1=
2(\bgamma+\bdelta)-1$, so equations (\ref{eq2}) and (\ref{eq3})
became \EQ\label{eq4}\bgamma+\delta \leq \kappa+\delta\leq
\gamma+\delta\leq 2(\bgamma+\bdelta)-1.\EN

If $\bdelta=0$ from equation (\ref{eq1}) we have $\delta=0$ and
$\beta=0$. From~equation~(\ref{eq4}) we can write $\bgamma\leq
\kappa\leq \gamma\leq 2\bgamma -1$. In this case, we can add that
$\kappa=\gamma$, since the $\gamma-\kappa$ vectors of order two
are the ones that are independent when we restrict them to the
quaternary part, but there is not quaternary part because
$\beta=0$.

If $\bdelta=1$ it is not possible to have two independent vectors
of order four in $\langle\codi_1,\codi_2\rangle$, so $\delta=1$
and from equation~(\ref{eq4}) we have $\bgamma \leq \kappa \leq
\gamma\leq 2\bgamma$.

If $\bdelta > 1$ then $2\bdelta=2(t-r)\leq 2^{t-1}-2^{r-1}=\beta$.
So, from equation~(\ref{eq1}) we obtain
$\delta\in\{\bdelta,\ldots,2\bdelta\}$ and
from~equation~(\ref{eq4}) we obtain $\bgamma \leq \kappa \leq
\gamma\leq 2\bgamma+2\bdelta-\delta-1$. \end{demo}

Recall that the parameters of an additive code can be computed
from the parameters of its additive dual code using
equations~(\ref{parameters}), so we can establish the following
theorem.

\begin{theo} \label{intersectionStructureBoundsAdditive} For any
two additive extended perfect codes $\codi_1$ and $\codi_2$ of
dual type ($\alpha,\beta;\bgamma,\bdelta$) with $\alpha \neq 0$,
the intersection code $\codi_1\cap\codi_2$ is of dual type
($\alpha,\beta;\gamma,\delta;\kappa$), where
$\gamma,\delta,\kappa$ satisfy the bounds given by
Proposition~\ref{lemmaBoundsAdditive}. \end{theo}

\begin{ex} \label{ExBeta4Z2Z4} For $t=3$ there are two isomorphic
additive extended perfect codes $\codi_1$ and $\codi_2$ given by
$\bdelta=0$ and $\bdelta=1$, so of dual types $(8,0;4,0)$ and
$(4,2;2,1)$, respectively. The code $\codi_1$ corresponds to an
extended binary Hamming code of length 8, so we have intersections
codes of dual type $(8,0;\gamma,0;\kappa)$ for any value
$\gamma=\kappa $ from 4 to 7 (see \cite{EV98} or
(\ref{EtzionVardyHamIntersection})). By an exhaustive search and
taking ${\cal H}_2$ as a parity-check matrix of $\codi_2$, the
possible intersection codes $\codi_2 \cap \pi(\codi_2)$, which
have parity-check matrices ${\cal H}_2 \parallel \pi({\cal H}_2)$,
are of dual type $(4,2;\gamma,\delta;\kappa)$, where $${\cal
H}_2=\left(\begin{array}{cccc|cc} 1&1&1&1& 2 & 2 \\ 0&0&1&1&0&2 \\
\hline 0&1&0&1&1&1 \end{array}\right) \quad \textrm{and} \quad
\begin{array}{cccc} \gamma & \delta & \kappa & \pi \\
\hline
2 & 1 & 2 & Id \\
3 & 1 & 2 & (1,2) \\
3 & 1 & 3 &  (1,3) \\
4 & 1 & 3 &  (1,2,3). \end{array}$$ \end{ex}

\begin{lemm} \label{lemmaInterStructure2betaAdditive} Let
$\codi_1$ and $\codi_2$ be additive extended perfect codes of dual
type $(\alpha,\beta;\gamma,\delta)$ with $\alpha \neq 0$,
$\alpha+2\beta=2^t$ and $t\geq 3$, such that $\codi_1 \cap
\codi_2$ is of dual type $(\alpha,\beta; i,j;k)$. Then, there
exist two additive extended perfect codes of dual type
$(2\alpha,2\beta; \gamma+1,\delta)$ with intersection codes of
dual type $(2\alpha,2\beta; i',j';k')$ for $(i',j',k')\in \{
(i+1,j,k+1), (i+2,j;k+1), (i+2,j;k+2) \}$. \end{lemm}

\begin{demo} Let ${\cal H}_1=( {\cal H}_{1,\alpha} | {\cal
H}_{1,\beta})$ and ${\cal H}_2=( {\cal H}_{2,\alpha} | {\cal
H}_{2,\beta})$ be parity-check matrices of the additive extended
perfect codes $\codi_1$ and $\codi_2$, respectively. The matrices
$$\left (\begin{array}{cc|cc}
0\ldots 0 & 1\ldots 1 & 0\ldots 0 & 2\ldots 2\\
{\cal H}_{1,\alpha}& {\cal H}_{1,\alpha} & {\cal H}_{1,\beta}&
{\cal H}_{1,\beta} \end{array} \right ) \  \textrm{and} \ \left
(\begin{array}{cc|cc}
0\ldots 0 & 1\ldots 1 & 0\ldots 0 & 2\ldots 2 \\
{\cal H}_{2,\alpha}& {\cal H}_{2,\alpha} & {\cal H}_{2,\beta}&
{\cal H}_{2,\beta}\end{array} \right )$$ are parity-check matrices
of additive extended perfect codes ${\cal D}_1$ and ${\cal D}_2$
of dual type ($2\alpha,2\beta;\gamma+1,\delta$). Using similar
arguments than in Lemma \ref{lemmaInterStructure2beta}, we can
take the permutations $\pi=Id$, $\pi=(1,\alpha+1)$ and
$\pi=(2\alpha+1,2\alpha+\beta+1)$, in order to obtain the
intersection codes ${\cal D}_1 \cap \pi({\cal D}_2)$ of dual type
($2\alpha,2\beta; i+1,j;k+1$), ($2\alpha,2\beta;i+2,j;k+2$) and
($2\alpha,2\beta;i+2,j;k+1$), respectively. \end{demo}

\begin{lemm} \label{lemmaInterStructure4betaAdditive} Let
$\codi_1$ and $\codi_2$ be additive extended perfect codes of dual
type $(\alpha,\beta;1,\delta)$ with $\alpha \neq 0$,
$\alpha+2\beta=2^t$ and $t\geq 3$, such that $\codi_1 \cap
\codi_2$ is of dual type $(\alpha,\beta; i,j;k)$. Then, there
exist two additive extended perfect codes of dual type
$(2\alpha,\alpha+4\beta; 1,\delta+1)$ with intersection codes of
dual type  $(2\alpha,\alpha+4\beta; i',j';k')$ for $ (i',j',k')
\in \{ (i,j+1;k), (i,j+2;k), (i+1,j+1;k), (i+1,j+1;k+1) \}$.
\end{lemm}

\begin{demo} Let ${\cal H}_1=( {\cal H}_{1,\alpha} | {\cal
H}_{1,\beta})$ and ${\cal H}_2=( {\cal H}_{2,\alpha} | {\cal
H}_{2,\beta})$ be parity-check matrices of the additive extended
perfect codes $\codi_1$ and $\codi_2$, respectively, such that
they contain the vector $(1 \ldots 1 | 2 \ldots 2)$ in the first
row.  The matrices $$\left (\begin{array}{cc|ccccc}
{\cal H}_{1,\alpha}& {\cal H}_{1,\alpha} & 2{\cal H}_{1,\alpha} & {\cal H}_{1,\beta} & {\cal H}_{1,\beta} & {\cal H}_{1,\beta} & {\cal H}_{1,\beta}\\
0\ldots 0 & 1\ldots 1  & 1\ldots 1 & 0\ldots 0 & 1\ldots 1  & 2\ldots 2 & 3\ldots 3 \\
\end{array} \right ) \ \textrm{and}$$
$$\left(\begin{array}{cc|ccccc}
{\cal H}_{2,\alpha}& {\cal H}_{2,\alpha} & 2{\cal H}_{2,\alpha} & {\cal H}_{2,\beta} & {\cal H}_{2,\beta} & {\cal H}_{2,\beta} & {\cal H}_{2,\beta}\\
0\ldots 0 & 1\ldots 1  & 1\ldots 1 &0\ldots 0 & 1\ldots 1& 2\ldots 2 & 3\ldots 3  \\
\end{array} \right )$$ are parity-check matrices of additive
extended perfect codes ${\cal D}_1$ and ${\cal D}_2$ of dual type
($2\alpha,\alpha+4\beta;1,\delta+1$). Notice that the first row in
these two matrices is again the vector $(1\ldots 1 | 2\ldots 2)$.
The codes $\codi_1$ and $\codi_2$ are of binary length
$\alpha+2\beta=2^t$, so the codes ${\cal D}_1$ and ${\cal D}_2$
are of binary length
$2\alpha+2(\alpha+4\beta)=4(\alpha+2\beta)=2^{t+2}$.

Using similar arguments than in Lemma
\ref{lemmaInterStructure4beta}, we can take the permutations
$\pi=Id$, $\pi=(2\alpha+1,3\alpha+\beta+1)$,
$\pi=(3\alpha+1,3\alpha+2\beta+1)$ and $\pi=(1,\alpha+1)$, in
order to obtain the intersection codes ${\cal D}_1 \cap \pi({\cal
D}_2)$ of dual type ($2\alpha,\alpha+4\beta;i,j+1;k$),
$(2\alpha,\alpha+4\beta; i,j+2;k)$, $(2\alpha,\alpha+4\beta;
i+1,j+1;k)$ and $(2\alpha,\alpha+4\beta; i+1,j+1;k+1)$,
respectively. \end{demo}

\begin{lemm} \label{lemmaInterStructureAdditiveG1} For all $m\geq
2$ there exist two additive extended perfect codes $\codi_1$ and
$\codi_2$ of dual type $(2^m,2^{2m-1}-2^{m-1};1,m)$, such that
$\codi_1 \cap \codi_2$ is an additive extended perfect code of
dual type $(2^m,2^{2m-1}-2^{m-1};\gamma,\delta;\kappa)$, where
$\gamma,\delta,\kappa$ are any values $$\delta \in \{m,\ldots,2m
\} \quad \textrm{and} \quad 1 \leq \kappa \leq \gamma \leq
2m-\delta+1$$ (so any values given by Theorem
\ref{intersectionStructureBoundsAdditive}). \end{lemm}

\begin{demo} By Lemma \ref{lemmaInterStructure4betaAdditive} and
the same argument than in Lemma
\ref{lemmaInterStructureBetaGeneral}, the result follows. We only
need to prove it for $m=2$. Let $\codi_1$ be the perfect code of
dual type $(4,6;1,2)$ with parity-check matrix $${\cal
H}_1=\left(\begin{array}{cccc|cccccc} 1&1&1&1&2&2&2&2&2&2 \\ 0&0&1&1&1&1&1&1&0&2 \\
\hline 0&1&0&1&0&1&2&3&1&1 \end{array}\right).$$ The possible
intersection codes $\codi_1 \cap \pi(\codi_1)$ are of dual type
$(4,6;\gamma,\delta;\kappa)$, where $$\begin{array}{cccc} \gamma &
\delta & \kappa & \pi \\ \hline
1&2&1& Id \\
2&2&1& (5,7) \\
2&2&2& (1,2)\\
3&2&1& (1,3)(6,9)(8,10)\\
3&2&2& (1,3)(5,7) \\
3&2&3& (1,2,3) \\
1&3&1& (5,6)\\
2&3&1& (5,6)(9,10)\\
2&3&2& (1,2)(6,9)\\
1&4&1& (5,6)(7,9). \end{array}$$ \end{demo}

\begin{theo} \label{structureadditive} For all $t\geq 4$ and any
$\bgamma, \bdelta$, such that $\bgamma+2\bdelta=t+1$, there exist
two additive extended perfect codes $\codi_1$ and $\codi_2$ of
dual type ($\alpha,\beta;\bgamma,\bdelta$), with $\alpha\neq 0$
and $\alpha+2\beta=2^t$, such that $\codi_1 \cap \codi_2$ is an
additive code of dual type ($\alpha,\beta;\gamma,\delta;\kappa$),
where   $$\left \{\begin{array}{lll} \mbox{if } \bdelta =0 &
\mbox{ then }& \delta=0 \mbox{ and } \bgamma\leq \kappa=\gamma\leq
2\bgamma
-1,\\
\mbox{if } \bdelta=1 & \mbox{ then } & \delta=1 \mbox{ and
}\bgamma \leq
\kappa \leq \gamma\leq 2\bgamma,\\
\mbox{if }\bdelta >1 & \mbox{ then } & \delta\in
\{\bdelta,\ldots,2\bdelta\} \mbox{ and }\bgamma \leq \kappa \leq
\gamma \leq 2\bgamma+2\bdelta-\delta-1. \end{array}\right.$$
except for codes of dual type $(8,4;3,1)$ for which the
intersection code of dual type $(8,4;6,1;3)$ does not exist.
\end{theo}

\begin{demo} By Theorem \ref{ExistPerfectZ2Z4LinearCodes}, we know
that for each $t\geq 4$ there are $\lfloor (t+2)/2 \rfloor$
non-isomorphic additive extended perfect codes of binary length
$\alpha+2\beta=2^t$. Specifically, for each
$\bdelta\in\{0,\ldots,\lfloor t/2 \rfloor \}$ there exists one of
dual type
($2^{t-\bdelta},2^{t-1}-2^{t-\bdelta-1};t+1-2\bdelta,\bdelta$).
Notice that when $\alpha+2\beta=2^{2m-1}$ we have $\bdelta \in
\{0,\ldots,m-1\}$ and when $\alpha+2\beta=2^{2m}$ we have $\bdelta
\in \{0,\ldots,m\}$.

For $t=4$, we have three non-isomorphic additive extended perfect
codes given by $\bdelta=0, 1$ and 2. For $\bdelta=0$, the code
corresponds to an extended binary Hamming code and for these codes
the result was proved in \cite{EV98} (see
(\ref{EtzionVardyHamIntersection})). For $\bdelta=1$, using Lemma
\ref{lemmaInterStructure2betaAdditive} and the codes constructed
in Example \ref{ExBeta4Z2Z4}, we can obtain intersection codes of
all different dual types except for $(8,4;6,1;3)$ and
$(8,4;6,1;6)$. By an exhaustive search, the intersection code of
dual type $(8,4;6,1;3)$ does not exist. However, taking the
permutation $\pi=(1,8,7,6,5,4,3)$ the intersection code $\codi_1
\cap \pi(\codi_1)$ is of dual type $(8,4;6,1;6)$, where $\codi_1$
is the perfect code of dual type $(8,4;3,1)$ constructed using
Lemma \ref{lemmaInterStructure2betaAdditive}.

Like in Theorem~\ref{theoremgapsquaternary}, in order to use
induction in this proof, since for $t=4$ and $\bdelta=1$ the
intersection code of dual type $(8,4;6,1;3)$ does not exist, we
need to show the existence of the intersection code of dual type
$(16,8;8,1;4)$ for $t=5$ and $\bdelta=1$. Again taking the
permutation $\pi=(1, 13, 10, 5)(2, 14, 9, 6)(3, 16, 12, 8)\\(4,15,
11, 7)(17, 22, 18, 20, 24, 21)(19, 23)$ the intersection code
$\codi_2 \cap \pi(\codi_2)$ is of dual type $(16,8;8,1;4)$, where
$\codi_2$ is the perfect code of dual type $(16,8;4,1)$
constructed using Lemma \ref{lemmaInterStructure2betaAdditive}.
Finally, by Lemmas \ref{lemmaInterStructure2betaAdditive} and
\ref{lemmaInterStructureAdditiveG1} and using similar arguments
than in Theorem \ref{theoremgapsquaternary}, the result follows.
 \end{demo}

Next theorem describes the intersection numbers for the additive
extended perfect codes with $\alpha\not=0$.

\begin{theo} \label{boundsAdditive} For any $t\geq 3$ and any
 additive extended perfect codes $\codi_1$ and $\codi_2$ of dual type
($\alpha,\beta;\gamma,\delta$), with $\alpha\neq 0$ and
$\alpha+2\beta = 2^t$, it is true that $$
\left\{\begin{array}{lll} \mbox{ if } \delta=1 & \mbox{ then } &
  2^{\alpha+2\beta-2t} \leq \eta(\codi_1, \codi_2) \leq
2^{\alpha+2\beta-t-1},\\
\mbox{ if } \delta\not=1 & \mbox{ then }&
  2^{\alpha+2\beta-2t-1} \leq \eta(\codi_1, \codi_2) \leq
2^{\alpha+2\beta-t-1}.\end{array}\right.$$ \end{theo} \begin{demo}
It is straightforward from Proposition \ref{lemmaBoundsAdditive}
and using the same argument than in Theorem
\ref{intercestionNumberQuaternary}. \end{demo}

Next result is to point out that the bounds in the previous
theorem are tight. Moreover, we show that there exist such codes
for any possible intersection number between these bounds. It is
easy and straightforward to settle this from Theorem
\ref{structureadditive}.

\begin{theo} \label{intersectionNumbersAdditive} For any $t\geq 3$
there exist two additive extended perfect codes ${\cal C}_1$ and
${\cal C}_2$ of dual type ($\alpha,\beta;\gamma,\delta$), with
$\alpha\neq 0$ and $\alpha+2\beta=2^t$, such that $\eta({\cal
C}_1, {\cal C}_2) = 2^{\alpha+2\beta-l}$, where $l$ is any value
such that \begin{displaymath} \left\{ \begin{array}{lll}
\mbox{ if } \delta=1 & \mbox{ then } & l \in \{t+1,\ldots,2t\}, \\
\mbox{ if } \delta \neq 1 & \mbox{ then } & l\in \{t+1,\ldots,2t+1\}. \\
\end{array}\right. \end{displaymath} \end{theo}

Finally, to end this section, notice that the usual binary perfect
codes of length $2^t-1$ are obtained from the extended ones by
puncturing one coordinate. The additive perfect codes can also be
constructed taking the parity-check matrix of an additive extended
perfect code, deleting the row with ones in the binary part and
twos in the quaternary part and also deleting one column in the
binary part.

Given any two additive perfect codes $\codi_1$ and $\codi_2$ of
dual type ($\alpha,\beta;\bgamma,\bdelta$) with $\alpha \neq 0$
and $\alpha+2\beta=2^t-1$, we can construct the extended codes
$\codi_1'$ and $\codi_2'$, respectively. The codes $\codi_1'$ and
$\codi_2'$ are of dual type ($\alpha+1,\beta;\bgamma+1,\bdelta$).

Using the theorems that we established before for additive
extended perfect codes, it is easy to get the same results for the
intersection codes $\codi_1\cap\codi_2$. We can summarize these
results with the following theorem:

\begin{theo} \label{nonextended} For any two additive perfect
codes $\codi_1$ and $\codi_2$ of dual type
($\alpha,\beta;\bgamma,\bdelta$) with $\alpha \neq 0$ and
$\alpha+2\beta=2^t-1$, the intersection code $\codi_1\cap\codi_2$
is of dual type ($\alpha,\beta;\gamma,\delta;\kappa$), where
$\gamma,\delta,\kappa$ satisfy: $$\left\{\begin{array}{lll}
\mbox{if } \bdelta=1 & \mbox{ then } & \delta=1 \mbox{ and }
\bgamma \leq \kappa \leq
\gamma \leq 2\bgamma+1, \\

\mbox{if } \bdelta\not=1 & \mbox{ then } & \delta\in
\{\bdelta,\ldots,2\bdelta\}  \mbox{ and } \bgamma \leq \kappa \leq
\gamma \leq 2\bgamma+2\bdelta-\delta. \end{array} \right.$$

\noindent For all $t\geq4$ and any values for
$\gamma,\delta,\kappa$ between these bounds there exist additive
perfect codes the intersection of which attains the prescribed
values, except for codes of dual type ($7,4;2,1$) for which the
intersection code of dual type ($7,4;5,1;2$) does not exist.
\end{theo}

This last theorem includes the binary Hamming codes (when
$\delta=0$ and, so, $\beta=0$), and we can see it as a
generalization of the solution of the intersection problem for
binary Hamming codes given by Etzion and Vardy (see
(\ref{EtzionVardyHamIntersection})).

\section{Conclusions}

In this paper we continue studying  the intersection problem for
codes initiated in \cite{EV98} (where the authors proposed to find
the intersection numbers for binary perfect codes) and
investigated in \cite{BYE97,AHS05,AHS06,PV06,SL06}.

Given two additive perfect codes we compute not only the
possibilities for the intersection number, but also the abelian
group structure of this intersection. We settle the problem for
non-extended and extended additive perfect codes, which means that
we solved the problem for perfect $\Z_4$-linear and $\add$-linear
codes.

There are still some interesting problems about this topic as, for
example, the problem of finding the abelian group structure of the
intersection for additive Hadamard codes, so the dual codes of the
additive extended perfect codes studied in this paper. Although we
know the relationship between the parameters of a given additive
code and its additive dual, and it would be easy to find
appropriate lower and upper bounds for the intersection structure,
it is not straightforward to construct all the codes which have
the desired parameters. This last point needs further research and
currently we are working on that.

\end{document}